\documentstyle[aps,preprint,float,prc]{revtex}
\tighten
\input{boxedeps.tex}
\SetRokickiEPSFSpecial
\HideDisplacementBoxes

\newcommand{\GG}{{\gamma\gamma}}
\newcommand{\EPEM}{e^+e^-}

\newcommand{\BE}{\begin{equation}}
\newcommand{\EE}{\end{equation}}

\begin{document}
\draft

\title{Production of Low Mass Electron Pairs Due to the 
Photon-Photon Mechanism in Central Collisions}
\author{Kai Hencken and Dirk Trautmann}
\address{
Institut f\"ur theoretische Physik der Universit\"at Basel,
Klingelbergstrasse 82, 4056 Basel, Switzerland
}
\author{Gerhard Baur}
\address{
Institut f\"ur Kernphysik (Theorie), Forschungszentrum J\"ulich, 
52425 J\"ulich, Germany
}

\date{August 23, 1999}

\maketitle

\begin{abstract}
We calculate the probability for dilepton production in central
relativistic heavy ion collisions due to the $\GG$ mechanism. This is
a potential background to more interesting mechanisms. We find that
this mechanism is negligible in the CERES experiments. Generally, the
contribution due to this mechanism is small in the central region,
while it can be large for small invariant masses and forward or
backward rapidities. A simple formula based on the equivalent photon
approximation and applications to a possible scenario at RHIC are also
given.
\end{abstract}


\section{Introduction}
\begin{figure}[htb]
\begin{center}
\ForceHeight{4cm}
\BoxedEPSF{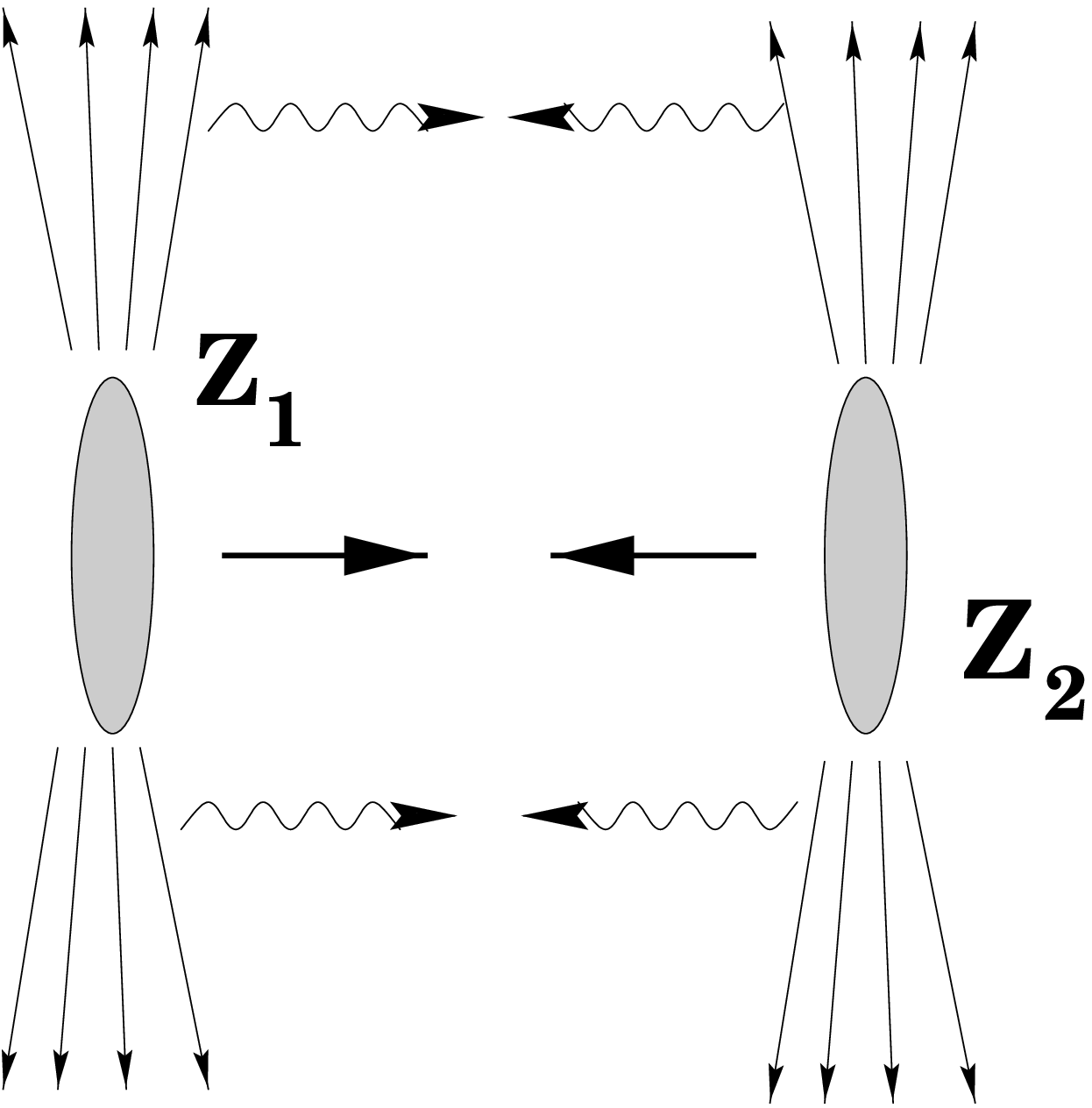}
\end{center}
\caption{The strong electromagnetic fields surrounding the heavy ions
are also present in central collisions ($b=0$) and will lead to pair
production processes due to the $\gamma\gamma$ fusion mechanism.}
\label{fig_collb0}
\end{figure}

In peripheral relativistic heavy ion collisions there are huge effects
due to the strong electromagnetic fields, while strong interactions
among the colliding ions can be virtually neglected, due to their
short range \cite{BaurHT98,HenckenSTB99}. In central collisions, on
the other hand, the strong interactions between the colliding ions
completely overwhelm the effects due to the electromagnetic
interaction.  It is the purpose of this paper to study the effects due
to the interactions of the very strong (coherent) electromagnetic
fields in the central collisions, see Fig.~(\ref{fig_collb0}). The
modification of the fields during the collision due to the strong
interactions will be neglected for simplicity. We are mainly
interested in the order of magnitude of the effects. Not unexpectedly,
the production of strongly interacting particles (like $\pi_0$) via
the $\GG$-mechanism is negligible. On the other hand, the photon-photon
mechanism could play a role in the production of $\EPEM$-pairs (and
also muon pairs). In recent experiments \cite{Agakichiev95} low mass
electron pairs are measured in p-Be, p-Au, and S-Au collisions. Even
heavier systems like Au-Pb have recently been studied
\cite{Lenkeit98}. For the proton induced interactions, the spectra are
explained essentially by electron pairs from hadronic decays, whereas
for the heavier systems an enhancement is obtained, which is suggested
to come from two-pion annihilation. In this paper we are going to
calculate the contribution of the $\GG$-mechanism to the electron
pairs.

In section~\ref{sec_theory} the general theoretical framework is
presented. Calculations within the framework of lowest order QED
\cite{HenckenTB94} are given in section~\ref{sec_calc} and compared to
order of magnitude estimates using the equivalent photon
approximation. In these calculations we have used the same kinematical
constraints as are used in the CERES experiment. We also present
calculations for a possible scenario at RHIC.  Our conclusions are
given in section~\ref{sec_concl}.

\section{Collisions of quasireal photons in central heavy ion interactions}
\label{sec_theory}
\subsection{Lowest order QED approach}
\label{ssec_qed}

The vector potential of the combined electromagnetic field of two
heavy ions, which travel with constant four velocities $u^{(1,2)}$ and
with an impact parameter $b$ between them is given by
\begin{eqnarray}
A_\mu (q) &=& - 2 \pi e \frac{1}{q^2} \biggl[ Z_1 F_1(q^2)u_\mu^{(1)} 
\delta (q u^{(1)}) \exp(i q b/2)  \nonumber\\
&& +  Z_2 F_2(q^2) u_\mu^{(2)} \delta (q u^{(2)}) \exp(-i q b/2)
\biggr],
\label{eq_fields}
\end{eqnarray}
where the four-velocities are given by
$u^{(1)}=(\gamma,0,0,\gamma\beta)$ and
$u^{(2)}=(\gamma,0,0,-\gamma\beta)$ \cite{HenckenTB95b}. In the
following we restrict ourselves to central collisions with $b\approx
0$. The nuclear charges and form factors are denoted by $Z_i$ and
$F_i(q^2)$, respectively ($i=1,2$).  It should be noted that due to
the strong interaction between the heavy ions, the nuclear charge
distributions (which give rise to the electromagnetic field) will
change; both stopping, as well as, a spreading out will take place. Such
effects were considered in \cite{BaurB93b}.  In addition there is also
pair production from the stopping itself \cite{LippertTGS91}. This
pair production due to the bremsstrahlung will strongly depend on the
actual stopping. We will neglect these stopping effects here. The
collision of the two electromagnetic fields (equivalent photons) can
then give rise to all kind of final states $f$, like $e^+ e^-$ pairs,
$\pi_0$, etc..

The matrixelement for $\EPEM$ pair production is given by
\begin{eqnarray}
  M &=& - i e^2 \bar u(p_-) \nonumber\\
&&\times    \int \frac{d^4p}{(2 \pi )^4}
    {\not\!\!A}(p_- - p) \frac{{\not\!p} + m}{p^2 - m^2}
    {\not\!\!A}(p_+ + p) 
    \;v(p_+),
\end{eqnarray}
where $\bar u(p_-)$ and $v(p_+)$ are the Dirac spinors describing the
produced electron and positron, respectively.  For impact parameter
$b$ equal to zero, this can be calculated completely analytically as
was done in \cite{HenckenTB94}. The integration over all final states
was then done with a MC integration \cite{Lepage78,Lepage80} in the
program code BORNZERO. The form factor of a pointlike charge
distribution ($F(q)=1$) contains unrealistically high Fourier
components, which can lead to spurious effects. Already the total
probability diverges logarithmically with $b\rightarrow 0$ in this
case, see \cite{HenckenTB95b}. In order to get realistic results a
nuclear form factor is needed in the calculations. We use here either
a monopole form factor
\BE
  F_{monopole}(q^2) = \frac{\Lambda^2}{\Lambda^2-q^2},
\EE
where $\Lambda$ is chosen to reproduce the correct rms radius of the
nucleus, as well as, one which is the sum of two monopole form
factors. We use them as a check on the sensitivity of our results on
the detailed form of the form factor. Compared to a ``realistic'' form
factor, the monopole form factor does not fall off rapidly enough for
large $q$. On the other hand this means, that our result is an upper
bound.

\subsection{Equivalent Photon approximation (EPA)}
\label{ssec_epa}
The impact parameter dependent equivalent photon number
$N(\omega,\rho)$ for a monopole form factor is given as
\cite{BaurF91,HenckenTB94}
\begin{eqnarray}
N(\omega,\rho) &=& \frac{Z^2 \alpha}{\pi^2} \Biggl| 
\frac{\omega}{\gamma} K_1\left(\frac{\omega}{\gamma} \rho\right)
\nonumber\\ &&
-\left[ \frac{\omega^2}{\gamma^2} +\Lambda^2 \right]^{1/2}
K_1\left( \left[ \frac{\omega^2}{\gamma^2} +\Lambda^2 \right]^{1/2}
\rho \right) \Biggr|^2.
\label{eq_nepa}
\end{eqnarray}
A simple approximation neglecting the contribution from the nuclear
interior is
\BE
N(\omega,\rho)= \frac{Z^2 \alpha}{\pi^2 \rho^2}, \quad
R<\rho<\gamma/\omega
\label{eq_nwapprox}
\EE
and 0 otherwise. The radius of the ion is given by
$R=1.2$~fm~$A^{1/3}$.  With this expression a simple formula for the
product of the equivalent photon numbers integrated over $\rho$ can be
given:
\BE
\int d^2\rho N(\omega_1,\rho) N(\omega_2,\rho)
\approx \frac{Z_1^2 Z_2^2 \alpha^2}{\pi^3 R^2} 
\left[1 - a^2 e^{2 |Y|}\right],
\EE
where $Y=1/2 \ln(\omega_1/\omega_2)$ and $a=R_{>} M /2\gamma<1$, with $R_{>}$ 
the larger one of the two nuclear radii and $M$ the invariant mass of
the produced system.

In the EPA the transverse momentum of electron and positron are the
same but in opposite directions ($\vec p_t(e^+)=-\vec p_t(e^-)$). We
have the following expression for the probability to emit an $e^+e^-$
pair with rapidities $y_+$ and $y_-$ and transverse momentum
$p_t(e^+)$
\BE
\frac{d^4P}{d^2p_t dy_+ dy_-} = \int d^2\rho N(\omega_1,\rho)
N(\omega_2,\rho) \frac{1}{\pi} \frac{d\sigma}{d\hat t},
\label{eq_epadiff}
\EE
where $d\sigma/d\hat t$ is the cross section for the subprocess
$\gamma+\gamma\rightarrow e^+ + e^-$, see, e.g., \cite{PeskinS95}.

Similarly the probability to produce a final state with invariant mass
$M$ and rapidity $Y$ is given by
\BE
\frac{d^2P}{dMdY} = \frac{Z_1^2 Z_2^2 \alpha^2}{\pi^3 R^2}
\left[ 1 - a^2 e^{2|Y|} \right] \frac{2 \sigma_{TT}(m^2)}{m},
\EE
where $\sigma_{TT}$ is the corresponding $\gamma+\gamma\rightarrow f$
photon-photon fusion cross section.

From this equation it can be easily seen that, e.g., $\pi_0$
production is very small:
\BE
\frac{dP}{dy} = \frac{ 8 Z_1^2 Z_2^2 \alpha^2 \Gamma_{\gamma\gamma}(\pi^0)}
{ \pi R^2 M_{\pi}^3} \left[ 1 - a^2 e^{2 |Y|} \right].
\EE
For $Y=0$ we find (for $\gamma$ large enough so that $a\ll1$) that the
pion production probability $dP/dY\approx 1.3 \times 10^{-5}$ for
Pb-Au or Au-Au collisions.

Since the $\pi^0$ is the lightest hadron, it is also the hadronic state
which will be produced with the highest probability. Therefore the
production of hadronic final states by the electromagnetic fields will
be negligible compared to the hadronic production.  We can therefore
concentrate in the following only on $\EPEM$ pairs, similar
calculations are of course also possible for muon or $\tau$ pairs. As
there was some concern regarding $\EPEM$ pairs coming from the
electromagnetic production in connection with the experiments
\cite{Hayano90}, we calculate them here exactly.

\section{Numerical results and Comparison with Experiment}
\label{sec_calc}

We want to compare our results directly with the measurements of the
CERES experiment \cite{Agakichiev95,Lenkeit98} and also want to make a
prediction for a possible scenario at RHIC \cite{Rapp99}. For
comparison with CERES we have used the same kinematical restriction
($p_t> 200$MeV for both S-Au and Pb-Au collisions). The
CERES experiment also has a restriction on the relative angles between
electron and positron ($\Theta_{ee}>35$mrad). This was not taken into
account, as tests have shown that --- apart from the very small
invariant masses --- this condition was always fulfilled.

\begin{figure}[htb]
\begin{center}
\ForceHeight{8cm}
\BoxedEPSF{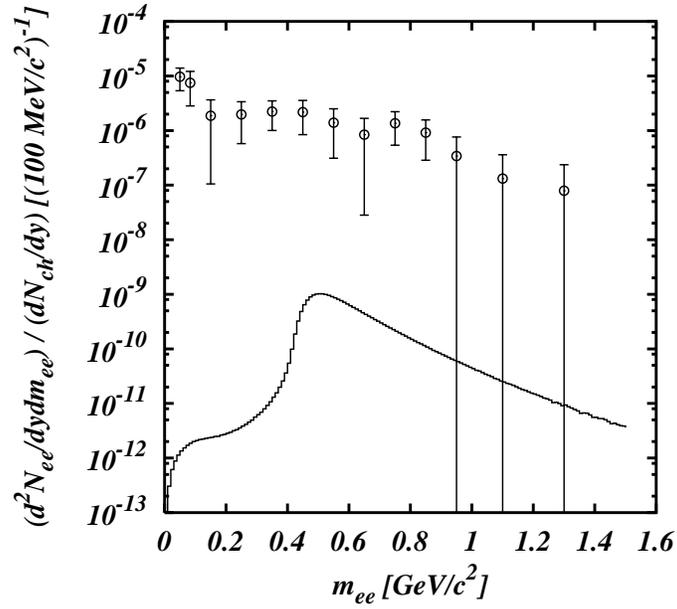}
\end{center}
\caption{$d^2N_{ee}/dydm_{ee}$ is shown for the electromagnetic
electron-positron pair production for central S-Au collisions at CERES 
($p_{t,min}>200$MeV$/c$); also shown are the experimental results of
the CERES experiment \protect\cite{Agakichiev95}.}
\label{fig_ceressau}
\end{figure}

\begin{figure}[htb]
\begin{center}
\ForceHeight{8cm}
\BoxedEPSF{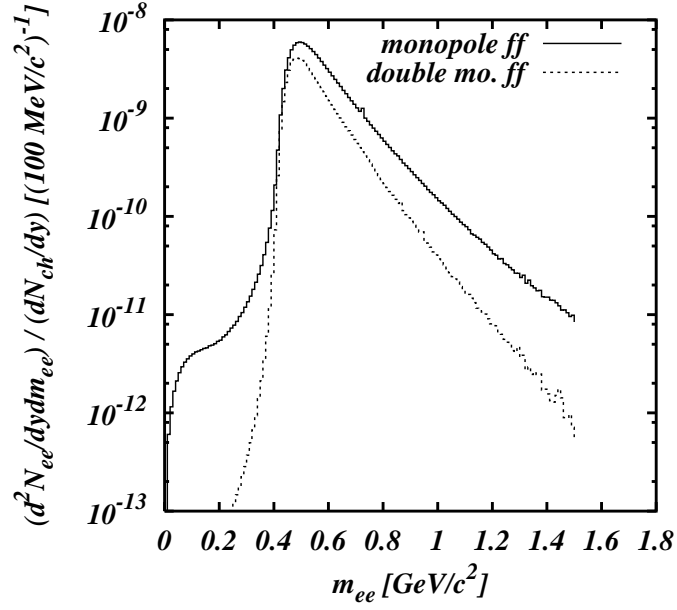}
\end{center}
\caption{$d^2N_{ee}/dydm_{ee}$ is shown for the electromagnetic
electron-positron pair production for central Pb-Au collisions at CERES
($p_{t,min}>200$MeV$/c$) and for two different form factors; see text
for details.}
\label{fig_cerespbau}
\end{figure}

\begin{figure}[htb]
\begin{center}
\ForceHeight{8cm}
\BoxedEPSF{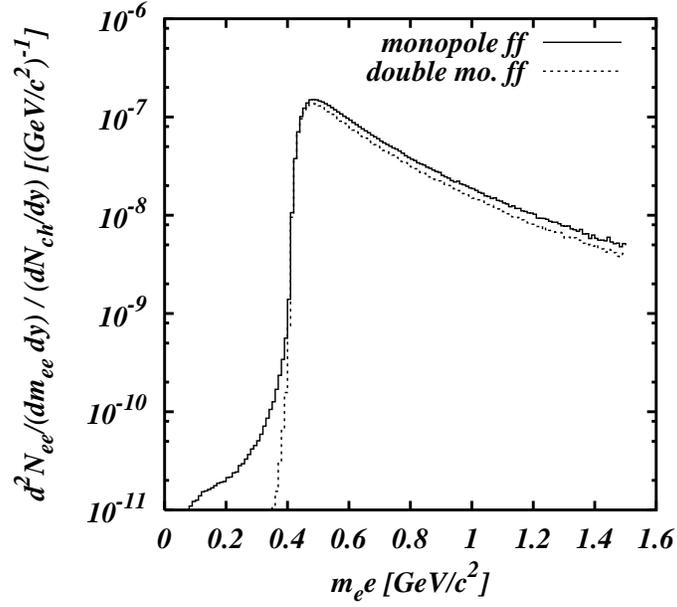}
\end{center}
\caption{Same as Fig.~\protect\ref{fig_cerespbau} for central Au-Au
collisions at RHIC; see text for details.}
\label{fig_rhicauau}
\end{figure}

The results of our calculations are shown in
Figs.~\ref{fig_ceressau}--\ref{fig_rhicauau}. We compare directly
$(d^2N_{ee}/dy dm_{ee})/(dN_{ch}/dy)$, using $dN_{ch}/dy=125$ for S-Au
collisions, $dN_{ch}/dy=235$ (the average of two experimental runs of
CERES for Pb-Au collisions.) We show the quantity $d^2N/dy dm_{ee}$,
which is defined to be the differential probability with respect to
the $y$ of either the electron or the positron and integrated over the
allowed $y$ range of the other ($2.1<y<2.65$ in the lab frame,
corresponding to $-0.93<y<-0.38$ in the ``center of velocity'' frame),
following \cite{Agakichiev95}.

For the RHIC scenario, see Fig.~\ref{fig_rhicauau}, we have used the
conditions as discussed in \cite{Rapp99}: $p_t>200$~MeV,
$-0.35<y<0.35$. $dN_{ch}/dy$ was assumed to be 1100.

We note the following points: The results of our calculations are well
below the experimental results of \cite{Agakichiev95} and
\cite{Lenkeit98}. Since the cross-section for the $\GG$ mechanism
scales with $Z_1^2 Z_2^2$ we expect the strongest effects for the
heaviest systems. For the p-Be and p-Au case at 450 GeV
\cite{Agakichiev95}, the $\GG$ mechanism is only a small fraction, as
expected. We do not need to show this. Even for the heaviest systems
\cite{Lenkeit98}, the effects are well below the
measurements. Therefore we can safely conclude that the $\GG$-mechanism
plays only an insignificant role for the experimental results of
\cite{Agakichiev95,Lenkeit98}.

Calculations with both the single and the double monopole form factor
are shown for Pb-Au and Au-Au collisions. The results for the CERES
experiment are found to be sensitive to the form of the factor both at
low and high invariant masses, those for RHIC only for the
low-invariant mass region ($m_{ee}< 2 p_t$). The sensitivity in this
range is rather easy to understand: In order to be in a low invariant
mass state, even though the momenta is at least $p_t$, electron and
positron cannot be produced in opposite transverse directions, as is
assumed in the equivalent photon approximation. In the EPA the
invariant mass of the lepton pair is given by
\BE
m_{ee}=2 p_\perp \cosh\left[1/2 \ (y_+ - y_-)\right].
\EE
Therefore the sum of their momenta is larger than the typical range of
the form factor (about 50~MeV for Pb or Au). This makes them sensitive
to the high $q$ part of the form factor.

The discrepancy at higher invariant masses at the CERES experiment
comes from a different reasoning: In order to produce a pair with, for
example, $m_{ee}=1$~GeV and with $Y=0$, one needs photons with energy
about $\omega=500 MeV$ (estimated from the equivalent photon
approximation). The kinematics of this process relates this with the
virtuality of the photon:
\BE
Q^2 = q_{\perp}^2 + \left(\omega/\gamma\right)^2.
\EE
With $\gamma\approx 10$ the virtuality $Q^2$ of the photon is then at
least 50~MeV at CERES, but only at least 7~MeV at RHIC. Therefore at
CERES one is again sensitive to the form factor at large virtuality.

For an order of magnitude estimate, we can use the EPA. We use
Eq.~(\ref{eq_epadiff}) together with the equivalent photon number $N$
of Eq.~(\ref{eq_nepa}) and the exact expression for $d\sigma/d\hat t$,
see, e.g., \cite{LandauL86}, and integrate over $d^2p_t$.

One can find also a simple analytic formula for the probability to
produce dileptons in the central region. We can neglect the square of
the electron mass as compared to the values of the Mandelstam
variables $\hat s$ and $\hat t$ of the subprocess $\gamma + \gamma
\rightarrow e^+ + e^-$ and find
\BE
\frac{d\sigma}{d\hat t} = \frac{2 \pi \alpha^2}{\hat s^2} \left
( \frac{ \hat s + \hat t}{\hat t} + \frac{ \hat t}{\hat s + \hat t}
\right).
\label{eq_sigt}
\EE
For $y_+ \approx y_-$ we have $\hat t= - \hat s/2$ and we find
\BE
\frac{dN_{ee}}{dm_{ee}} \approx (\Delta y)^2 2
\frac{Z_1^2 Z_2^2 \alpha^4}{\pi^2 R^2 m^3}.
\label{eq_analytic}
\EE

\begin{figure}[htb]
\begin{center}
\ForceHeight{8cm}
\BoxedEPSF{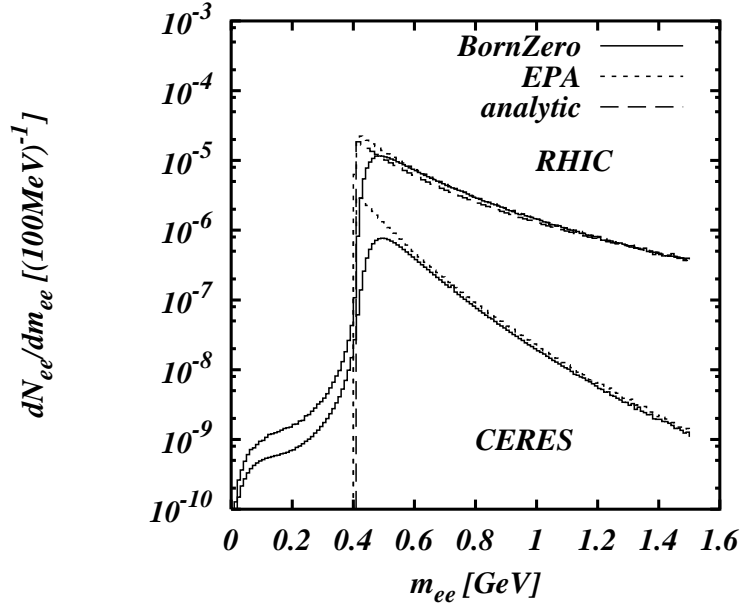}
\end{center}
\caption{Different approximations to the pair production probability
are compared. Shown are the results of the full calculation
(BornZero), the full EPA calculation (EPA, Eq.~(\ref{eq_epadiff}) with
$N$ of Eq.~(\ref{eq_nepa}) and the analytic expression
of Eq.(\protect\ref{eq_analytic}) (analytic).}
\label{fig_epa}
\end{figure}

In Figure~\ref{fig_epa} the different approaches are compared both for
Au-Au collisions at RHIC ($-0.35<y<0.35$, $\Delta y=0.7$) and Pb-Au
collisions at CERES ($2.1<y<2.65$, $\Delta y=0.55$). One finds that
EPA agrees quite well at invariant masses above $2 p_t$ with the exact
calculation. The analytic expression given above is in good agreement
with the results at RHIC, but as explained above, is not valid for the
conditions at CERES, as the parameter $a$ is larger than 1 (EPA would
predict in this case ($a>1$) a value of zero!).
Equation~(\ref{eq_sigt}) can be rewritten in terms of a differential
cross section $d\sigma/d\Omega$, which results in a
$\left[1+\cos^2(\theta)\right]/\sin^2(\theta)$ dependence on the
scattering angle in the center of mass system. From this one sees that
the importance of the $\gamma\gamma$ mechanism, when one goes to more
forward or backward angles (larger rapidities), increases as compared
to the other mechanisms for dilepton production, see also
\cite{BaurB93b}.

\section{Conclusions}
\label{sec_concl}

A comparison with the measured $\EPEM$ pairs in central collisions at
CERES shows, that pairs produced electromagnetically from the strong
electric fields, are only a small background.  We think that this
calculation has, for the first time, convincingly shown that the
$\gamma\gamma$ pairs are negligible. This is mainly due to the
experimental conditions on large transverse momenta and large
invariant masses. The majority of the electron-positron pairs is
produced with a invariant mass of the order of $1-10\ m_e$ with rather
small $p_t$. Also the photon-photon mechanism is strongly peaked at
very forward angles and therefore $\GG$-pairs will show up prominently
under such kinematical conditions

We extended our calculations also to RHIC energies. In this energy
regime, $\GG$-pairs are produced with invariant masses up to several
GeV. Therefore the $\GG$ mechanism can be potentially dangerous in
this case. Again, our calculations show that this is only a small
effect in the central region, where we compare our calculations with
the predictions given in \cite{Rapp99}. In any case, with the present
calculations the contributions due to the $\gamma\gamma$ mechanism are
well under control. They are reassuringly small in the central region,
where one is looking for the effects of hot and dense matter.


%
%

\end{document}